\documentclass[12pt,a4paper]{article}

\usepackage{amsmath}
\usepackage{amssymb}
\usepackage[nosort]{cite}
\usepackage{graphicx}

\newcommand{\publj}{j}
\newcommand{\puble}{e}
\let\publ=\puble
\newcommand{\befigpar}[1]{\ifx\publ\puble \begin{#1}[ht]\else \begin{#1}[hp]\fi}

\newcommand{\hypref}[2]{\ifx\href\asklfhas #2\else\href{#1}{#2}\fi}

\newcommand{\secref}[1]{Sec.~\ref{#1}}
\newcommand{\figref}[1]{Fig.~\ref{#1}}
\newcommand{\tabref}[1]{Tab.~\ref{#1}}

\newcommand{\sfrac}[2]{{\textstyle\frac{#1}{#2}}}

\newcommand{\conjugate}{{\ast}}

\newcommand{\newop}[2]{\def#1{\mathop{\mathrm{#2}}\nolimits}}
\newop{\artanh}{artanh}
\newop{\diag}{diag}
\newop{\Re}{Re}
\newop{\Im}{Im}

\newcommand{\I}{i}
\newcommand{\decay}{f}

\newcommand{\Reals}{{\mathbb{R}}}

\newcommand{\bigbrk}[1]{\bigl(#1\bigr)}

\newcommand{\coeffv}[2]{v_{#1}^{(#2)}}

\newcommand{\cbeta}[2]{\beta_{#1}^{(#2)}}

\newcommand{\Lagr}{{\mathcal{L}}}

\newcommand{\deltaph}{\lambda}

\newcommand{\tsum}{\mathop{\textstyle\sum}\nolimits}

\newcommand{\MeV}{\,\mathrm{MeV}}
\newcommand{\GeV}{\,\mathrm{GeV}}

\begin{document}

\thispagestyle{empty} 

\hfill\hfill 

\bigskip\bigskip\bigskip

\begin{center}
\Large\bf\boldmath  $S$-wave Meson-Meson Scattering\\ 
from Unitarized $U(3)$ Chiral Lagrangians%
\end{center}

\bigskip\bigskip

\begin{center}
{\large N. Beisert\footnote{email: nbeisert@physik.tu-muenchen.de}
 and  B. Borasoy\footnote{email: borasoy@physik.tu-muenchen.de}}

\bigskip\bigskip

\hypref{http://www.ph.tum.de/}{Physik-Department},
Technische Universit{\"a}t M{\"u}nchen\\
D-85747 Garching, Germany
\end{center}

\bigskip\bigskip\bigskip\bigskip

\begin{abstract}
An investigation of the  
$s$-wave channels in meson-meson scattering is performed within a $U(3)$ 
chiral unitary approach. 
Our calculations are based on a chiral
effective Lagrangian which includes the $\eta'$ as an explicit degree of
freedom and incorporates important features of the underlying QCD
Lagrangian such as the axial $U(1)$ anomaly.
We employ a coupled channel Bethe-Salpeter equation 
to generate poles from composed states of two pseudoscalar mesons.
Our results are compared with experimental phase shifts up to $1.5\GeV$
and effects of the $\eta'$ within this scheme are discussed.
\end{abstract}\bigskip

\begin{center}
\begin{tabular}{ll}
\textbf{PACS:}&12.39.Fe, 11.10.St, 11.80.Gw, 14.40.-n\\[6pt]
\text {Keywords:}& 
Chiral Lagrangians, 
quasi-bound states, 
coupled channels, 
\\&
$U(3)$ chiral perturbation theory, $\eta'$.
\end{tabular}
\end{center}

\newpage
\section{Introduction}

The $SU(3)_L \times SU(3)_R$ chiral symmetry of QCD is spontaneously broken
down to $SU(3)_V$ giving rise to eight pseudoscalar Goldstone bosons, the pions, kaons, and the eta.
At low energies the interactions among these Goldstone bosons 
are described well in chiral perturbation theory (ChPT) which is the effective
field theory of QCD. The Green functions are ordered in powers of 
the small meson masses and momenta, such that they are organized as Taylor expansions.
This systematic perturbative chiral expansion is limited to the low energy region. 
At higher energies, the accuracy of
the chiral series decreases, until convergence finally fails and becomes useless. 
One reason for the failure of convergence is, e.g., the 
exchange of resonances between the mesons in scattering processes.
The resonances appear as poles in the scattering amplitude
and cannot be generated to any order in a plain series expansion.
Nevertheless it has been shown that, when combined with 
non-perturbative methods such as Lippmann-Schwinger equations (LSE)
which are employed in such a way as to ensure unitarity, the chiral Lagrangian
 is able to reproduce a number of observed resonances both in the purely
mesonic sector and  under the inclusion
of baryons, see e.g.
\cite{Kaiser:1995eg,Kaiser:1997js,Oller:1997ti,Oller:1999hw,Nieves:2000bx,CaroRamon:2000jf,Bass:2001np}.
Within these approaches effective coupled channel potentials are derived
from the chiral meson Lagrangian and iterated in Lippmann-Schwinger equations,  
or in the relativistic case Bethe-Salpeter equations (BSE).
(For simplicity we will not distinguish between the two.)
The BSE generates dynamically quasi-bound states of the mesons and baryons and 
accounts for the exchange of resonances without including them explicitly.
The usefulness of this approach lies in the fact that from a small set of
parameters a large variety of data can be explained.

In the purely mesonic sector Oller and Oset have used the BSE to probe
the system of two interacting mesons.
Employing the lowest order $SU(3)$ chiral Lagrangian they were able to 
generate a number of scalar resonances at around $1\GeV$,
which could be identified with the observed resonances
$f_0(980)$ and $a_0(980)$ \cite{Oller:1997ti}.
Furthermore, the resulting scattering cross sections 
were matched in good agreement with experimental data. 
By considering  fourth order ChPT in a subsequent work \cite{Oller:1999hw} the results were 
extended to account for further resonances below $1\GeV$,
e.g. the lowest-lying vector mesons $\rho$, $K^\ast$. 
The authors find agreement with data at energies up to $\sqrt{s} \simeq 1.2\GeV$ .
(Similar results are obtained in a fully relativistic $SU(2)$ ChPT approach
\cite{Nieves:2000bx}.)


The $\eta'(958)$, on the other hand, cannot be generated in coupled channel approaches 
by these two-meson states
due to its pseudoscalar nature. In fact,
the $\eta'$ meson is considered to be the singlet counterpart of the octet of Goldstone bosons
$(\pi,K, \eta)$. The extra mass of the $\eta'$ is due to the axial $U(1)$ anomaly
which prevents it from being a Goldstone boson. 
In the large $N_c$ limit the axial $U(1)$ anomaly vanishes
yielding nine Goldstone bosons. The $\eta'$ is then the ninth Goldstone boson with 
a mass comparable with the other mesons.
It is thus possible to combine the $\eta'$ meson with the octet of Goldstone bosons.
To this end, we will extend the chiral Lagrangian by including the $\eta'$ 
explicitly and without employing large $N_c$ rules.
We use the fourth order $U(3)$ 
chiral effective Lagrangian, see e.g. \cite{Kaiser:2000gs,Beisert:2001qb},
to evaluate the interaction kernel for the BSE.
All possible two-meson states are taken into account in a relativistic BSE approach 
to calculate the propagators of the pertinent quasi-bound states.
By restricting ourselves to conventional $SU(3)$ chiral Lagrangians and neglecting the $\eta'$
we are then able to study its effects in the coupled channel analysis
which may offer new insights into the importance of the axial anomaly.
The inclusion of the $\eta'$ may not only produce new resonances in the spectrum 
due to the appearance
of new channels, but can in principle also destroy the agreement with the well
established resonances of $SU(3)$ coupled channel analyses below
$1\GeV$.
Even if the channels which involve the $\eta'$ are below threshold 
and cannot contribute to physical processes directly, they 
can have effects on  channels with two Goldstone bosons via mixing.
Our investigation provides an important check whether a similar
agreement with experiment as in the $SU(3)$ case
can be obtained in the presence of the $\eta'$.

In the meson-baryon sector, the $SU(3)$ coupled channel formalism has already
been extended to include the $\eta'$, and meson-baryon scattering processes together
with photoproduction of $\eta$ and $\eta'$ on the proton have been investigated
\cite{Bass:2001np}. Within their approach the authors find substantial changes 
with respect to the original work in the $SU(3)$ sector \cite{Kaiser:1997js}.
Even after fitting the parameters in their approach, they were not able to achieve 
good agreement with experimental data in contrast to \cite{Kaiser:1997js}.
Sizeable
effects of the $\eta'$ were also observed in the processes in which the $\eta'$
is not an external particle, but contributes via virtual $\eta'$-baryon states
\cite{Stefan}. It is hence worthwile investigating whether the inclusion
of the $\eta'$ in the purely mesonic sector destroys the good agreement
of previous investigations with experimental data and whether one needs to finetune
the parameters in order to reachieve agreement. It could well be that -- as in the case
of \cite{Bass:2001np} -- the inclusion of the $\eta'$ does not allow for an overall 
good description of the data, even if there is no significantly big branching ratio
to the $\eta'$ and a Goldstone boson for the resonances discussed in the present work.

The inclusion of the $\eta'$ furthermore allows for a consistent treatment of 
$\eta_8$-$\eta_0$ mixing. In the $SU(3)$ framework, 
the $\eta$ is treated as the octet state $\eta_8$ with its mass
being at its physical value $m_\eta = 547\MeV$, while some effects of the $\eta'$, 
after integrating it out from the theory,
are hidden in coupling constants of the effective Lagrangian at next-to-leading
order \cite{Ecker:1989te, Herrera-Siklody:1998cr}.
In \cite{Beisert:2001qb} it was shown that $\eta_8$-$\eta_0$ mixing does not follow
the usually assumed one-mixing-angle scheme, but must be 
parametrized in terms of two angles even at leading order, if large $N_c$
counting rules are not imposed. In order to account for this unusual behavior,
one needs to include the $\eta'$ field explicitly.

Two-meson systems consisting of an $\eta'$ and a Goldstone boson will
lead to contributions in meson-meson scattering, e.g., the $\pi \eta'$
decay mode of the $s$-wave resonance $a_0(1450)$ is seen by the Crystal Barrel 
experiment \cite{Abele:1998qd} and the experimentally well studied $f_0(1500)$
has an $\eta \eta'$ decay mode \cite{Amsler:1995bz}. In \cite{Bass:2001zs} the
possibility that the $J^{PC}= 1^{-+}$ exotics observed at BNL \cite{Thompson:1997bs,Ivanov:2001rv}
and CERN \cite{Abele:1998gn} may be resonances in $\eta \pi$ and $\eta' \pi$
scattering was investigated. The authors come to the conclusion that it is 
indeed possible to describe the appearance of exotics by means of a coupled 
channel treatment of the $\eta \pi$ and $\eta' \pi$ systems.
Within that work, however, it was not checked whether the inclusion of 
$\eta'$ channels destroys the overall agreement of the coupled channel
analysis for $p$-wave meson-meson scattering at energies below $1.2\GeV$.
Our investigation will shed some light on the importance of the $\eta'$ channels
within coupled channel approaches for $s$-wave resonances 
and can be extended to $p$-waves.
This may help to understand the role of the axial anomaly and gluons
in the structure of these resonances.

This work is organized as follows.
In \secref{sec:ResBSE} we introduce the kinematics and
solve the Bethe-Salpeter equation.
The identification of poles in the complex continuation of
the two-particle propagator is discussed in \secref{sec:ResProp}
and 
\secref{sec:ResResults} contains the results of the analysis.

\section{Bethe-Salpeter Equation}\label{sec:ResBSE}

\subsection{Kinematics}

\def\figone{
\befigpar{figure}
\phantom{}\hfill
\parbox{7cm}{\centering\includegraphics{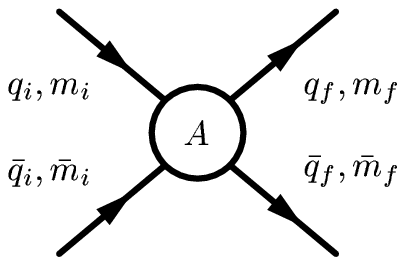}}
\hfill\hfill
\parbox{7cm}{\centering\includegraphics{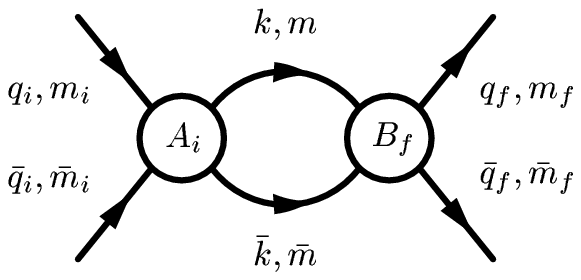}}
\hfill\phantom{}
\caption{Momenta and masses used in the four point scattering amplitudes and
chain links. The center-of-mass momentum is denoted by 
$p=q_i+\bar q_i=q_f+\bar q_f=k+\bar k$.}
\label{fig:ResAmpli}
\end{figure}}

\ifx\publ\puble\figone\fi

In this section we introduce our notation for the kinematics 
of the Bethe-Salpeter equation. 
We will work in the relativistic framework, restrict ourselves to $s$-waves
and put all momenta on-shell (see below).
The momenta and masses of a four-point scattering process are
given in \figref{fig:ResAmpli}.
A general scalar amplitude $A$ depends only on scalar combinations
of the momenta which can be expressed in terms of the Mandelstam variables.
The Mandelstam invariants $s$, $t$ and $u$ are defined as the center-of-mass 
energy squared $s=(q_i+\bar q_i)^2=p^2$, the momentum transfer
$t=(q_i-q_f)^2$ and the crossed momentum transfer
$u=(q_i-\bar q_f)^2$. 
The constraint $s+t+u=q_i^2+\bar q_i^2+q_f^2+\bar q_f^2=m_i^2+\bar m_i^2+m_f^2+\bar m_f^2$
allows one to neglect the combination $t+u$ in favor of $t-u$. 
The scalar amplitude can be written as $A(s,t-u)$.
Since we are only interested in scalar, i.e. $s$-wave or $l=0$, 
resonances, we must separate channels of different total angular momentum.
The amplitude $A$ in our approach is a fourth order polynomial in the momenta.
Hence, $A$ can be decomposed as 
\begin{equation}\label{eq:ResPartialWave}
A=\tsum\nolimits_l A_l J_l=A_s J_s+A_p J_p+A_d J_d ,
\end{equation}
where the partial wave operator $J_l$ 
is a polynomial of
degree $l$ in $t-u$.
The $J_l$ read
{\arraycolsep0pt\begin{eqnarray}
J_s=&&\mathrel{}1,\nonumber\\
J_p=&&\mathrel{}h_{\mu\nu}q_i^\mu q_f^\nu
=\frac{t-u}{4}+\frac{(q_i^2-\bar q_i^2)(q_f^2-\bar q_f^2)}{4s},
\nonumber\\
J_d=&&\mathrel{}D_{\mu\nu\,\rho\sigma}\, q^\mu_{i} q^\nu_{i} q_f^\rho q_f^\sigma 
=J_p^2-\frac{h_{\mu\nu}q_i^\mu q_i^\nu\,h_{\rho\sigma}q_f^\rho q_f^\sigma}{d-1},
\end{eqnarray}}%
and in $d=4$ space-time dimensions they are proportional 
to Legendre polynomials in the cosine of the scattering angle.
The metric $h$ of the $(d-1)$-dimensional space transverse to $p$ is given
by 
\begin{equation}
h_{\mu\nu}=-g_{\mu\nu}+p_\mu p_\nu/p^2.
\end{equation}
The partial wave operators $J_l$ can be given in terms of
spin projectors, e.g. the spin-$2$ projector
\begin{equation}
D^{\mu\nu}{}_{\rho\sigma}=
\sfrac{1}{2}h^\mu_\rho h^\nu_\sigma
+\sfrac{1}{2}h^\mu_\sigma h^\nu_\rho-h^{\mu\nu}h_{\rho\sigma}/(d-1).
\end{equation}
The spin projectors are totally symmetric in the upper (lower) indices, 
orthogonal to $p$ and 
have the property that every pair of upper (lower) indices is traceless;
they project to the spin-$n$ components of a general tensor of rank $n$. 
This formalism allows us to extract the $s$-wave part of the amplitude $A$.

\subsection{Bethe-Salpeter Equation}

\def\figtwo{
\befigpar{figure}
\[
\parbox{3cm}{\centering\includegraphics{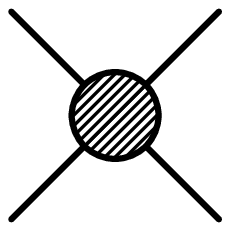}}
=
\parbox{3cm}{\centering\includegraphics{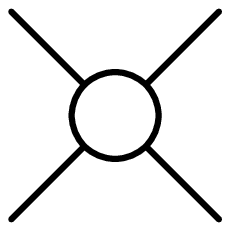}}
+
\parbox{5cm}{\centering\includegraphics{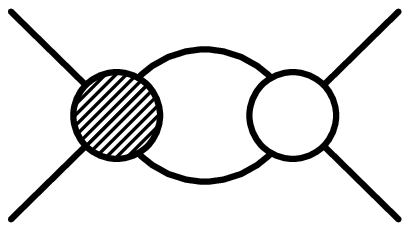}}
\]
\caption{Bethe-Salpeter equation for the interaction kernel $A$ (empty circle)
and solution $T$ (shaded circle).}
\label{fig:ResBSE}
\end{figure}}

\ifx\publ\puble\figtwo\fi

The Bethe-Salpeter equation for the two-particle propagator $T$ 
from a local interaction $A$ is given by 
\cite{Nieves:2000bx} 
\begin{equation}\label{eq:ResBSElong}
T(p,q_i,q_f)=A(p,q_i,q_f)+
\int\frac{i\,d^d k}{(2\pi)^d}\,\frac{T(p,q_i,k) A(p,k,q_f)}{(k^2-m^2)(\bar k^2-\bar m^2)}.
\end{equation}
or diagrammatically in \figref{fig:ResBSE}.
Combinatorial factors for two identical particles in the loop have to be taken 
into account, but we prefer to keep this form of the BSE and 
modify the amplitudes accordingly.
In that case $A$ must be the four-point amplitude from 
ChPT multiplied by $-\frac{1}{2}$ in order for
$T$ to be proportional to a bubble chain with
the correct factors from perturbation theory. The factor of
$\frac{1}{2}$ is the symmetry factor of two identical
particle multiplets in a loop and $-1$ stems from factors of $i$ in the
vertices and propagators.

We can further simplify the integral in the BSE \eqref{eq:ResBSElong}, since
we are only interested in the physically relevant piece of the solution $T$
with all momenta put on the mass shell.
The amplitude $A$ contains in general off-shell parts which deliver 
via the integral a contribution even to the on-shell part of the 
solution $T$.
However, these off-shell parts yield exclusively chiral logarithms 
which -- besides being numerically small -- can be absorbed
by redefining the regularization scale of the loop integral. 
Furthermore the off-shell parts are not uniquely defined in 
ChPT.
We will therefore set all the momenta in the amplitudes in \eqref{eq:ResBSElong}
on-shell.\footnote{This was also done 
in other work such as \cite{Oller:1997ti}. For a discussion of off-shell effects 
see \cite{Nieves:2000bx}.}
The BSE then simplifies to the arithmetic equation
\begin{equation}\label{eq:ResBSEtamed}
T =A+T G A,
\end{equation}
with $G$ being the scalar loop integral
in dimensional regularization
{\arraycolsep0pt\begin{eqnarray}
G_{m \bar{m}}(p^2) =&&\mathrel{}
\int\frac{i\,d^d k}{(2\pi)^d}\,\frac{1}{(k^2-m^2)( (k-p)^2-\bar m^2)}\nonumber \\
=&&  \mathrel{}
\frac{1}{16\pi^2}\bigg[-1+
\ln\frac{m \bar  m}{\mu^2}
  +\frac{m^2-\bar m^2}{p^2}\ln\frac{m}{\bar m}
\nonumber\\&&\qquad\qquad\quad\mathord{}
  -\frac{2\sqrt{\deltaph_{m\bar{m}}(p^2)}}{p^2}\artanh\frac{
  \sqrt{\deltaph_{m\bar{m}}(p^2)}}{(m+\bar m)^2-p^2}\bigg]
\nonumber\\
\deltaph_{m\bar{m}}(p^2)=&&\mathrel{}\big((m-\bar m)^2-p^2\big)\big((m+\bar m)^2-p^2\big).
\end{eqnarray}}%
The scattering matrix $S$ is given by
\begin{equation}\label{eq:ResUnitarity}
 S=1-i C T C\qquad\mbox{with }
 C^2=-2\Im  G
\end{equation}
which is unitary due to the identity $2 \Im T  = -T C^2 T^\conjugate$.
The phase of the unit complex number $S$ is parametrized by the 
scattering phase $\delta$ defined by $S=\exp(2i\delta)$.
For an ideal resonance $\delta$ increases by $180^\circ$. 

The generalization to $n$ coupled channels is achieved by promoting 
$A$, $T$, $G$, $C$ and $S$ to $n\times n$ matrices. 
The matrix $A$ contains the interaction kernels among the channels,
$G$ is diagonal with the loop integrals for the channels as elements.
The unitary scattering matrix $S$ can be given in terms of eigenvectors 
and eigenphases.

\section{Propagator in the Complex Plane}\label{sec:ResProp}

\subsection{Branch cuts}

Since the integral has physical significance only for real
values of $E=\sqrt{s}$, we will refer to the set of 
real $E$ as the `physical real axis'.
In the coupled channel analysis, however, 
we would like to identify structures on the real axis with poles of 
the analytic continuation in the lower half of the complex plane. The analytical 
continuation of $G_{m\bar m}(E^2)$ inherits several branch cuts from its
constituent functions and we use the following conventions.
The branch cut of $\sqrt{x}$ is just below the negative real axis, 
the branch cuts of $\artanh{x}$ are
below the negative real axis from $-1$ and above
the real axis from $+1$.
The resulting branch cut in $I$ is below the positive 
real axis of $E$ starting at the threshold point $m+\bar m$. 
This Riemann sheet is commonly referred to as the `physical' sheet.

\def\figthree{
\befigpar{figure}
\centering 
\phantom{}\hfill\includegraphics{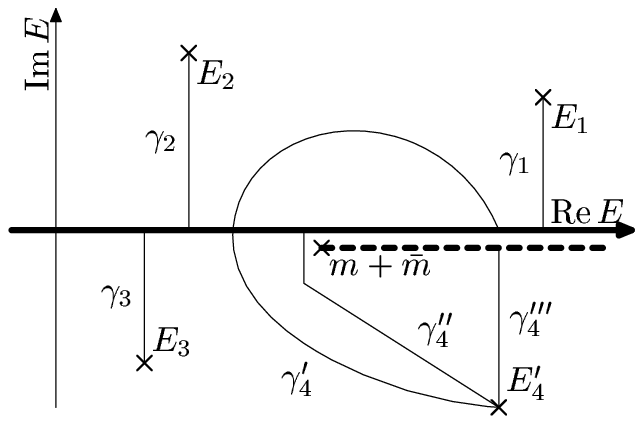}
\hfill\hfill\includegraphics{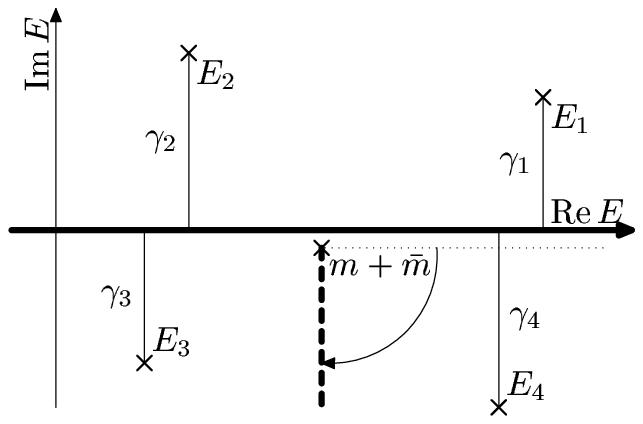}\hfill\phantom{}
\caption{Branch cuts in `physical' sheet (left) and our choice of Riemann
sheet (right). Shown
are paths from points $E_i$ to the physical real axis $E\in\Reals$.
The branching point $m+\bar m$ is moved slightly down for 
reasons of presentation.}
\label{fig:ResCuts}
\end{figure}}

\ifx\publ\puble\figthree\fi

Unfortunately, it is not well suited for finding 
physically relevant poles.
This can be seen as follows:
a relevant pole is a pole in the lower half of the complex plane and close to
the physical real axis which 
has a strong influence on observables. 
Points $E$ in the upper half of the 
complex plane or below threshold 
(e.g. $E_1$, $E_2$, $E_3$ in \figref{fig:ResCuts} left) 
can be connected with a straight vertical line $\gamma$
($\gamma_1$, $\gamma_2$, $\gamma_3$) to the physical real axis.
The length of $\gamma$ is $|\Im p|$ and
$E$ is as close to the physical real axis as possible.
Points $E$ in the lower half of the complex plane above threshold
(e.g. $E'_4$), on the other hand, are not close to the physical real axis,
because the length of the paths connecting $E$ to the physical real axis 
(such as $\gamma'_4$, $\gamma''_4$) exceeds $|\Im E|$.
The vertical path towards the real axis ($\gamma'''_4$) 
crosses the branch cut and ends in an unphysical real axis. 

As a matter of fact, the
region below the physical real axis and above
threshold turns out to be the most interesting of all, because here
 physically relevant poles occur.   
Therefore it is better to consider the `physical' sheet with the branch cut
rotated down by $90^\circ$ (\figref{fig:ResCuts} right). 
Every point in that rotated sheet is
as close as possible to the physical real axis and 
the loop integral function defined on the rotated sheet is 
\begin{equation}\label{eq:ResModInt}
G_{m\bar m}(E^2)\mapsto G_{m\bar m}(E^2)-\frac{\sqrt{-\deltaph_{m\bar m}(E^2)}}{8\pi E^2}
\,\theta\bigbrk{-\Im E}\,\theta\bigbrk{\Re(E-m-\bar m)},
\end{equation}
with $\theta(x)$ being the unit step function
with $\theta(0)=0$.
With this choice for the analytic continuation of the loop integral most poles that are relevant for
the physical real axis -- except those which
are related to cusps (cf. next section) -- are in the same sheet, whereas other
investigations have to take several sheets into account, in order to observe the physically
relevant poles.

\def\figfour{
\befigpar{figure}
\centering
\includegraphics{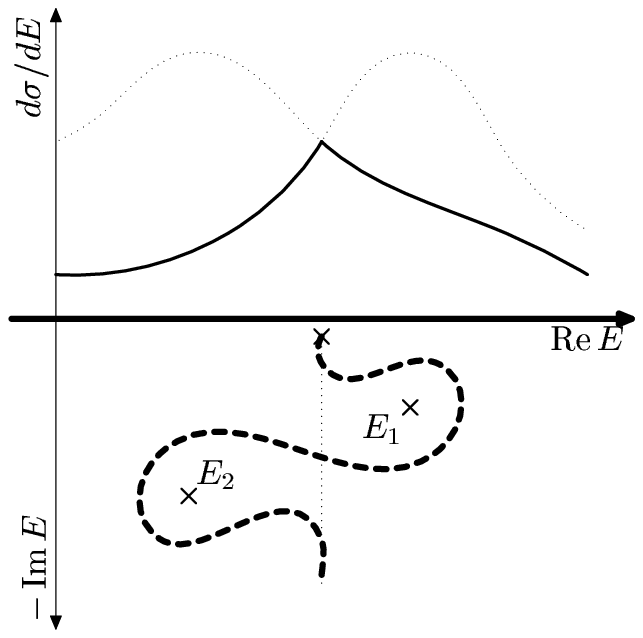}
\caption{A cusp at threshold and the corresponding pair of poles. 
The branch cut can be perturbed so that both poles reside on the same Riemann sheet.}
\label{fig:ResCusp}
\end{figure}}

\ifx\publ\puble\figfour\fi

\subsection{Cusps}

When the modified integral \eqref{eq:ResModInt} is used for all
channels in a coupled channel approach, most of the poles
can be seen simultaneously. Nevertheless, some relevant poles may still
be hidden behind a branch cut. 
This is the case if there are cusp resonances in the spectrum. 
Cusps are discontinuities of the derivative of
the full amplitude and they occur at the threshold energy of each channel. 
Cusp resonances can be generated by the configuration of 
poles as shown in \figref{fig:ResCusp}. In the rotated Riemann sheet no 
poles are seen, but there is one pole just behind the branch cut on either side. 
When the branch cut is moved around these two poles the
situation becomes clearer. The real axis below threshold is close to the
pole at $E_1$. The amplitude will therefore have a peak at $\Re E_1$,
but only the increase on the right of the resonance is below threshold and physical;
the peak itself lies on an unphysical real axis. Above threshold it is just
the opposite, the peak at $\Re E_2$ is hidden and only the decline on the left
is physical. The cusp resonance is therefore the interplay between two poles 
and a branch cut. 
It is even more instructive to consider both poles as manifestations of
the same pole disturbed by the branch cut singularity:
At a pole the denominator is zero. 
Going from one sheet to another corresponds to adding
some function to the propagator integral in the denominator.
Assuming this function to be small, the root will be shifted by a small amount
giving rise to a nearby pole on the new sheet. 
When $E_1$ and $E_2$ are considered one,
the cusp can be considered as a common resonance with the 
middle part removed by the branch cut.
In the scattering phase a cusp will correspond to an
increase of considerably less than $\pi$ as for usual resonances, because 
the sharp increase at the center is eliminated.

\section{Results and Comparison}\label{sec:ResResults}

We can now employ the Bethe-Salpeter equation
as presented in \secref{sec:ResBSE}, in order to fit
to scattering data of two 
pseudoscalar mesons. In $U(3)$ ChPT the fundamental pseudoscalar mesons are the 
Goldstone bosons $(\pi,K,\eta)$  with
their singlet counterpart, the $\eta'(958)$.
The potential $A$ is derived from the chiral $U(3)$ 
Lagrangian by
calculating the tree diagrams up to fourth chiral order
and taking $\eta$-$\eta'$ mixing into account where we employ the
two-mixing angle scheme as described in \cite{Beisert:2001qb}.
Tadpoles and crossed diagrams which are not included in our approach have been shown
to yield numerically small effects in scattering processes in the physical region
and can furthermore be partially absorbed by redefining chiral parameters.
We have therefore neglected both tadpoles and crossed diagrams and restrict ourselves
to the tree diagrams in the calculation of the potential $A$.

We work with the Lagrangian found in 
\cite{Beisert:2001qb,Herrera-Siklody:1997pm}
{\arraycolsep0pt\begin{eqnarray}
\Lagr=&&\mathrel{}
\sfrac{1}{4}f^2\langle \partial^\mu U^\dagger \partial_\mu U\rangle
+\sfrac{1}{2}f^2 \Re\langle U^\dagger \chi \rangle
+i\sfrac{1}{3}\sqrt{6}\,\coeffv{3}{1}(\ln\det U)\Im\langle U^\dagger\chi\rangle
+\sfrac{1}{12}f^2 m_0^2(\ln\det U)^2
\nonumber\\&&\mathord{}
+\cbeta{0}{0}\langle \partial^\mu U^\dagger \partial^\nu U
\partial_\mu U^\dagger \partial_\nu U\rangle
+\cbeta{3}{0}\langle \partial^\mu U^\dagger \partial_\mu U
\partial^\nu U^\dagger \partial_\nu U\rangle
\nonumber\\&&\mathord{}
+2\cbeta{5}{0}\Re\langle \partial^\mu U^\dagger \partial_\mu U U^\dagger \chi\rangle
+2\cbeta{8}{0}\Re\langle U^\dagger \chi U^\dagger \chi\rangle
+\ldots
\end{eqnarray}}%
where $\chi=2B\diag(\hat m,\hat m,m_s)$ is the quark mass matrix in the isospin limit
and
$m_0$ denotes the mass of the $\eta'$ in the chiral limit.
We have omitted all terms that are irrelevant or are not considered here, in particular
we have only taken the most relevant terms from the fourth chiral order Lagrangian
according to the OZI rule.
We replace the quark masses $\hat m$, $m_s$ and 
the constant $f$ by the fourth chiral order expressions (without loops) 
for the pion, kaon masses 
$M_\pi=138\MeV$, $M_K=496\MeV$
and the pion decay constant $F_\pi = 92.4\MeV$.
The interaction kernel $A$ for the Bethe-Salpeter equation are the
tree level amplitudes from the Lagrangian separated in 
angular momentum and isospin channels. 
As an example we state the $4\pi$ vertex with $J=0$, $I=2$, 
\[
A=\frac{s-2M_\pi^2}{2F_\pi^2}
-\frac{4\cbeta{0}{0}}{F_\pi^4}\bigbrk{s-2M_\pi^2}^2
-\frac{4\cbeta{3}{0}}{3F_\pi^4}\bigbrk{4M_\pi^4-2M_\pi^2 s+s^2}
+\frac{4\cbeta{5}{0} M_\pi^2 s}{F_\pi^4}
-\frac{16\cbeta{8}{0}M_\pi^4}{F_\pi^4}.
\]
For the particle propagators we use the standard
propagators for scalar particles with the physical masses of the particles.

In this section we present 
a fit to scattering data
from \cite{Mercer:1971kn} ($K\pi$),
\cite{Hoogland:1974cv,Losty:1974et} ($\pi\pi$, $I=2$)
and \cite{Ishida:1999xx}. 
The work \cite{Ishida:1999xx} is a collection 
of $\pi\pi$ scattering data from 
\cite{Hoogland:1974cv,Losty:1974et,Baton:1970pe,Baton:1970ma,Cason:1983qt,Estabrooks:1974vu,Protopopescu:1973sh,Rosselet:1977pu,Srinivasan:1975tj,
Cohen:1973yx,Colton:1971iw,Prukop:1974ia,Hoogland:1977kt,
Scharenguivel:1970yz,Hyams:1973zf,Zylbersztejn:1972kd,Grayer:1974cr}
and $K\pi$ scattering data from \cite{Aston:1988ir,Estabrooks:1978xe}.
Having replaced the quark masses and $\decay$ 
by the meson masses and $F_\pi$ as described above, the only free 
parameters left are the regularization scale $\mu$ in $G$
and the coupling constants of the effective Lagrangian. 

\subsection{SU(3) ChPT}

\def\tabone{
\befigpar{table}
\centering
\[\begin{array}{|ll|c|}\hline
           I&\mbox{resonance}& \mbox{pole}\\\hline
           0&f_0(400-1200)   &\phantom{0}448-          263\I   \\
           0&f_0(980)        &\phantom{0}983-\phantom{0}14\I   \\
\sfrac{1}{2}&K_0^*(900)      &\phantom{0}740-          246\I  \\
           1&a_0(980)^{<}    &          1081-\phantom{0}36\I   \\
           1&a_0(980)^{>}    &\phantom{0}791-\phantom{0}75\I   \\
\hline
\end{array}\]

\caption{Pole positions (MeV) of scalar channels from the  $SU(3)$ chiral Lagrangian 
and using the parameter set in \eqref{eq:ResBetaFit}.}
\label{tab:ResPoleSU3}
\end{table}}

\def\figfive{
\befigpar{figure}
\centering
\includegraphics[width=16cm]{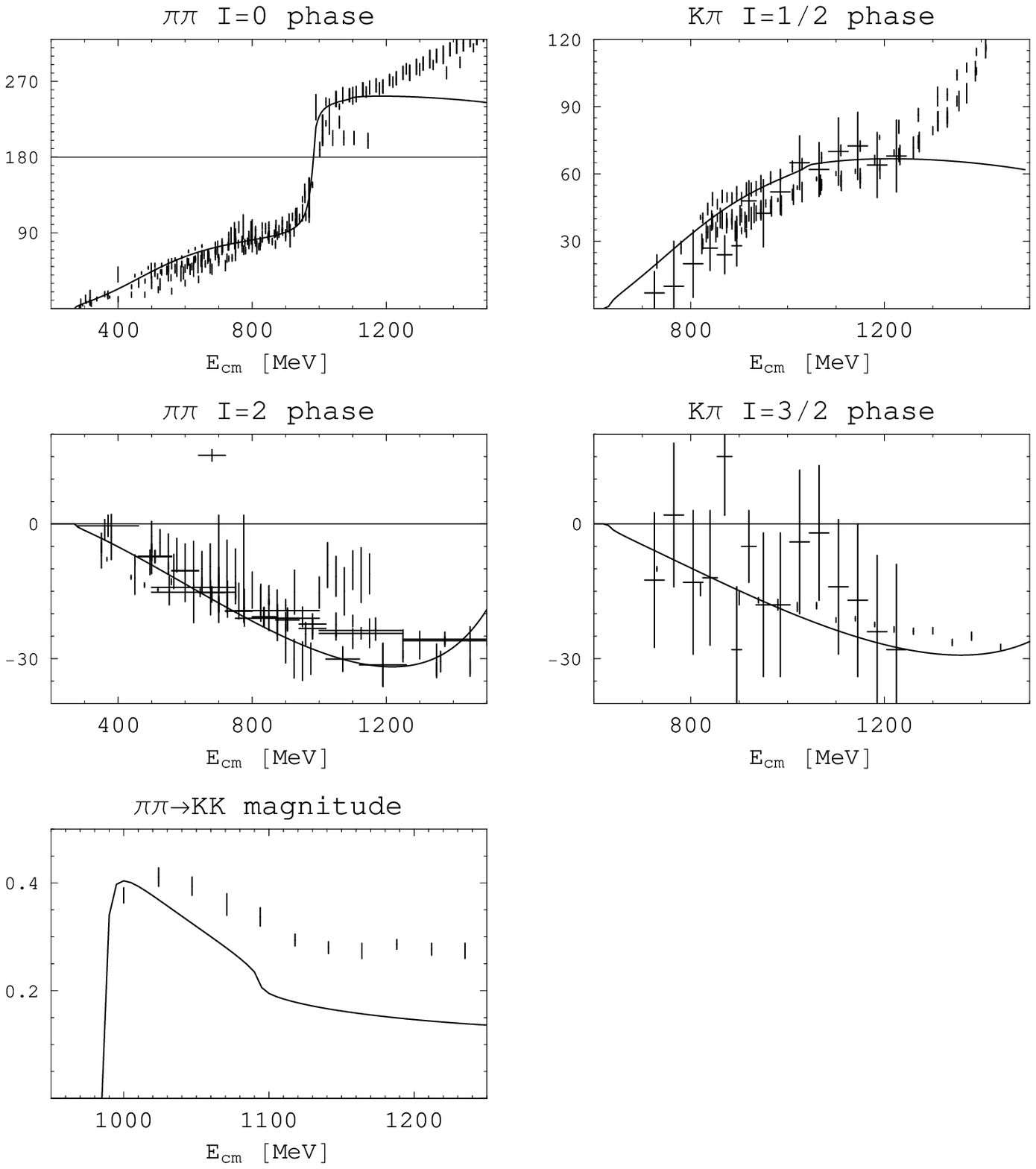}

\caption{Scattering data from $SU(3)$ chiral Lagrangians
and experimental data from 
\cite{Ishida:1999xx} (all plots),
\cite{Mercer:1971kn} ($K\pi$),
\cite{Hoogland:1974cv,Losty:1974et} ($\pi\pi$, $I=2$).
The $S$-matrix elements are parametrized as
$S=\sqrt{2\zeta} e^{2i\delta}$ where $\delta$ is the
phase and $\zeta$ the magnitude.
}
\label{fig:ResSU3pha}
\end{figure}}











We first restrict ourselves to the conventional chiral $SU(3)$ Lagrangian.
This is done by omitting all possible vertices which include the $\eta'$ field.
Later on we will proceed by including the $\eta'$ field explicitly. By comparison
we can then pin down the effects of the $\eta'$ within this approach.
The parameters entering in the pure $SU(3)$ case are besides the regularization scale $\mu$
only the known parameters $\beta_k$, $k=0, \ldots, 8$,
from $SU(3)$ ChPT. However, there is a slight difference in so far
as that we keep the low-energy constant (LEC) $\beta_0$ from the fourth order Lagrangian explicitly.
Usually this contact term is absorbed into other terms of the Lagrangian by
employing a Cayley-Hamilton matrix identity \cite{Herrera-Siklody:1997pm},
but for processes including the $\eta'$ it seems to be more convenient
to keep this term \cite{Beisert:2002ad}. The values of some of the LECs $\beta_k$, 
involved in the Cayley-Hamilton identity change accordingly. 
The LECs $\beta_1$, $\beta_2$, $\beta_4$, $\beta_6$, and $\beta_7$ are then 
compatible with zero within their phenomenologically
determined error bars and we neglect them as an approximation.
This estimate for the LECs has been proven to be quite successful in
\cite{Beisert:2001qb,Beisert:2002ad} 
and suggests that the important physics for the considered processes
is included in the remaining parameters $\beta_0$, $\beta_3$, $\beta_5$, and $\beta_8$.
The omission of the first parameters is also motivated by the observation that they
can be interpreted as OZI violating corrections of the latter ones.
Of course, an improved fit to data might be obtained by fine-tuning these suppressed
parameters, but no additional insight is gained and none of our conclusions change.
We therefore consider the reduced set of parameters 
$\cbeta{0}{0}, \cbeta{3}{0}, \cbeta{5}{0}$ and $\cbeta{8}{0}$ with 
their values given by
\begin{equation}
\label{eq:ResBetaFit}
\begin{array}{lll}
\cbeta{0}{0}=(0.6\pm 0.1)\times 10^{-3},&\quad&
\cbeta{3}{0}=(-0.5\pm 0.1)\times 10^{-3},
\\
\cbeta{5}{0}=(1.4\pm 0.2)\times 10^{-3},&\quad&
\cbeta{8}{0}=(0.2\pm 0.2)\times 10^{-3}.
\end{array}
\end{equation}
while neglecting the remaining LECs $\beta_k$.
Using this set of parameters we will compare the results for the phase shifts
with available experimental data.
The results presented in the paper have been obtained by employing the central values
for the LECs in Eq.~(\ref{eq:ResBetaFit}), but variations within the given ranges 
for the parameters do not
lead to substantial differences in the results.  
We furthermore restricted ourselves first
to a single scale parameter $\mu=1\GeV$ for all channels and energies.
However, with such a simplified choice the calculated phase in the $I=\sfrac{1}{2}$ channel 
turned out to be slightly above the data points.
This feature can also be seen in \cite{Oller:1999hw,Oller:1999ag}.
By lowering the scale down to $\mu=0.8\GeV$ in that particular channel we
are easily able to improve the fit.
This indicates that our approach neglects further contributions in the $I=\sfrac{1}{2}$ channel
which we mimic by fine-tuning the scale $\mu$.
The results are summarized in \tabref{tab:ResPoleSU3} and
\figref{fig:ResSU3pha}.

\ifx\publ\puble\tabone\fi

In the $I=0$ channel matching is remarkable up to energies 
around $E_{\mathrm{cm}}=1.2\GeV$, the 
linear increase from threshold to just below $1\GeV$ 
is due to the $\sigma$ or $f_0(400-1200)$ 
resonance and the $f_0(980)$ resonance can be clearly seen
by a sudden phase shift of $180^\circ$. These resonances are
associated with poles at 
$(448 - 263i)\MeV$ and $(983-14i)\MeV$.
This is in reasonable agreement with recent results for light scalar mesons obtained
from Dalitz plot analyses of charm decays in the Fermilab experiment E791 \cite{Bediaga:2002ei}.
By analyzing the decay $D^+ \rightarrow \pi^- \pi^+ \pi^+$ \cite{Aitala:2000xu} strong evidence of the
$\sigma$ was found with a mass of $478 \pm 17\MeV$ and a width of $324 \pm 21\MeV$
which corresponds to twice the imaginary part of the pole position.
From the analysis of the $D_s^+ \rightarrow \pi^- \pi^+ \pi^+$ 
decay \cite{Aitala:2000xt} the mass and the width of
the $f_0(980)$ were remeasured to be $975 \pm 3\MeV$ and $44 \pm 4\MeV$,
respectively.

Our results start deviating from the experimental phase shifts at around
$E_{\mathrm{cm}}=1.2\GeV$, however, this is not very surprising since higher particle 
effects which are omitted in this scheme, in particular the $4 \pi$ channel, 
will become important at these energies \cite{Kaminski:1997gc}. 
In the $I=\sfrac{1}{2}$ channel a broad resonance,
the $\kappa$ or $K_0^\ast(900)$, can be seen
extending from threshold to about $1\GeV$, which is related 
to a pole at 
$(740-246i)\MeV$.
Again, we have good agreement with data
up to energies of $E_{\mathrm{cm}}=1.3\GeV$.
In the $I=1$ channel a sharp increase just below $1\GeV$ is 
due to the $a_0(980)$ resonance, which manifests as a cusp in
the scattering amplitude.
A possible cusp interpretation of the $a_0$ has already been given in
\cite{Uehara:2002nv}.
In our analysis it corresponds to poles
at 
$(1081-36i)\MeV$ and $(791-75i)\MeV$
(see the discussion about cusps in \secref{sec:ResProp}).
These poles are both hidden on our standard Riemann sheet,
the first pole lies on the Riemann sheet corresponding to
the physical region between the branching points
of $\pi\eta$ at $682\MeV$ and $KK$ at $988\MeV$
and the second one corresponding to the Riemann sheet above the $KK$ branching point.
Therefore, the $a_0$ appears as a resonance with its central part
cut away and the phase shift is less than $180^\circ$.
For the $I=\sfrac{3}{2}$ and $I=2$ channels reasonable agreement with
experiment is achieved  for center-of mass energies up to $E_{\mathrm{cm}}=1.5\GeV$
and no significant increase of the phase shifts is observed.

\subsection{U(3) ChPT}

\def\tabtwo{
\befigpar{table}
\[\begin{array}{|ll|c|}\hline
           I&\mbox{resonance}& \mbox{pole}\\\hline
           0&f_0(400-1200)   &\phantom{0}459-          233\I   \\
           0&f_0(980)        &\phantom{0}994-\phantom{0}10\I   \\
\sfrac{1}{2}&K_0^*(900)      &\phantom{0}737-          248\I  \\
           1&a_0(980)^{<}    &          1061-\phantom{0}55\I   \\
           1&a_0(980)^{>}    &\phantom{0}761-\phantom{0}62\I   \\
\hline
\end{array}\]

\caption{Pole positions (MeV) of scalar channels from the  $U(3)$ chiral Lagrangian 
and using the parameter set in \eqref{eq:ResBetaFit}.}
\label{tab:ResPoleU3}
\end{table}}

\def\figsix{
\befigpar{figure}
\centering
\includegraphics[width=16cm]{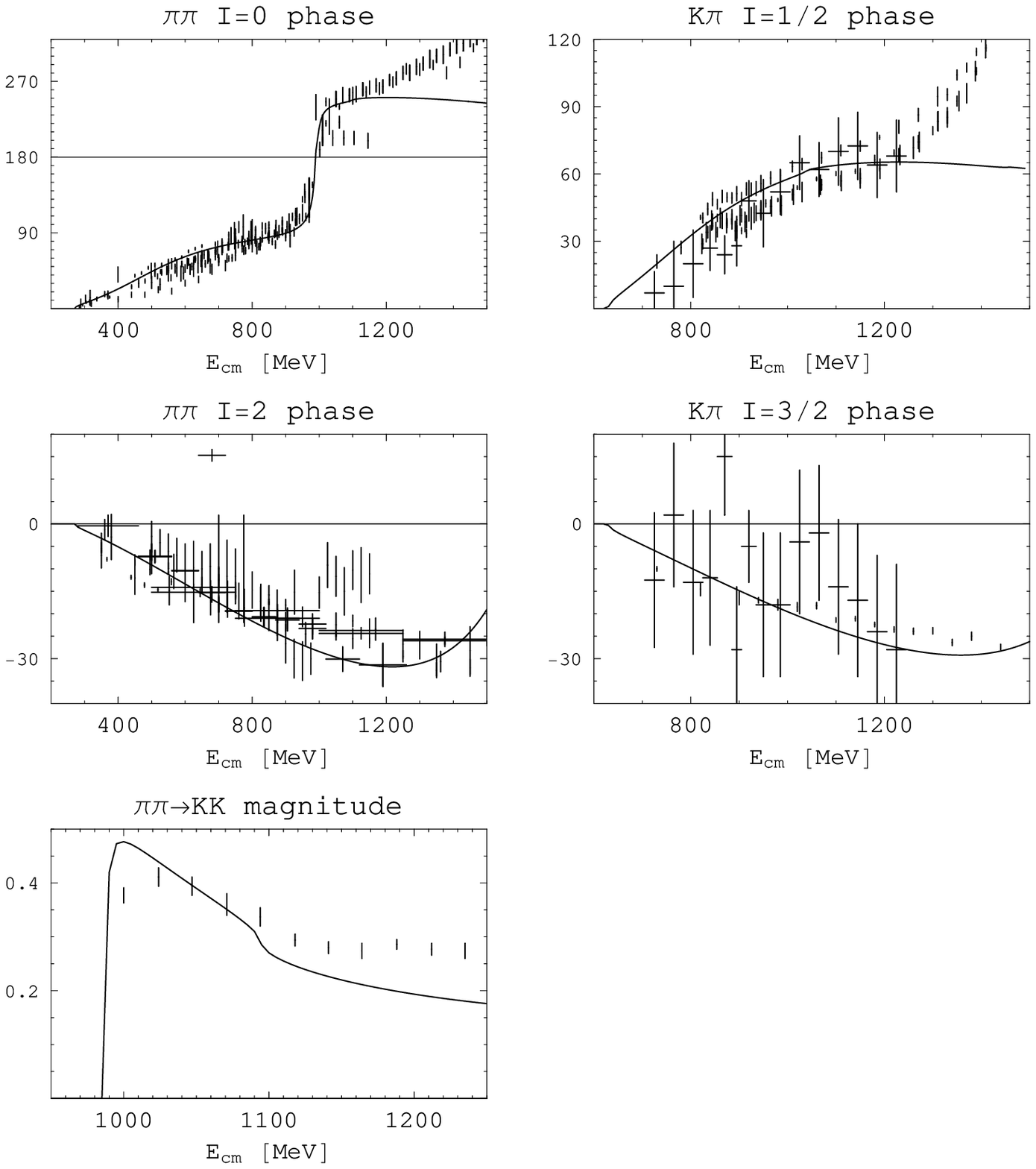}

\caption{Scattering data from $U(3)$ chiral Lagrangians
with experimental data points from 
\cite{Ishida:1999xx} (all plots),
\cite{Mercer:1971kn} ($K\pi$),
\cite{Hoogland:1974cv,Losty:1974et} ($\pi\pi$, $I=2$).
The $S$-matrix elements are parametrized as
$S=\sqrt{2\zeta} e^{2i\delta}$ where $\delta$ is the
phase and $\zeta$ the magnitude.
}
\label{fig:ResU3pha}
\end{figure}}

We now extend the chiral Lagrangian to its $U(3)$ form by including 
the $\eta'$ explicitly.
In order to compare the results with the pure $SU(3)$ analysis we make the
{\it same} choice for the parameters \eqref{eq:ResBetaFit}, while 
the additional LECs of the $U(3)$ Lagrangian
are set to zero. This approximation yields good results and we refrain from
performing a better fit to existing data by fine-tuning the new couplings.
The resulting scattering phases are shown in \figref{fig:ResU3pha} 
and the pole positions are given in \tabref{tab:ResPoleU3}.
We note that up to $1.5\GeV$ 
there are no considerable differences to $SU(3)$ ChPT.
In particular, the inclusion of the $\eta'$ channels does not yield
any new resonances in the considered energy range up to $1.5\GeV$. 

\ifx\publ\puble\tabtwo\fi

This is a non-trivial observation, since the inclusion of $\eta'$ channels
might have disturbed the agreement with experimental data of the pure
$SU(3)$ case via coupling between the channels as has been observed in the
meson-baryon sector \cite{Bass:2001np}. In the purely mesonic sector, on the other hand,
the effects of the 
$\eta'$ decouple to a large extent from the interactions of the Goldstone bosons.
Remarkably, the results are insensitive to
the value of $\tilde v_{2}^{(1)}=\sfrac{1}{4}f^2-\sfrac{1}{2}\sqrt{6}v_{3}^{(1)}$ which is 
mainly responsible for $\eta\eta'$ mixing \cite{Beisert:2001qb}. 
Variation of its value from $\tilde v_{2}^{(1)}=\sfrac{1}{4} f^2$ with omitted
$1/N_c$-suppressed piece $v_{3}^{(1)}$ down to $\tilde v_{2}^{(1)}=0$ for suppressed mixing
does not alter our results considerably.

The similarity of the $SU(3)$ and $U(3)$ results with the same set of parameters
depends on a Cayley-Hamilton identity which can be utilized 
to absorb the parameter $\beta_{0}$ by some of the other LECs.
In the $SU(3)$ case this identitiy involves the parameters $\beta_{0},\ldots,\beta_{3}$,
whereas for $U(3)$ the additional parameters $\beta_{13},\ldots,\beta_{16}$
are included, see \cite{Beisert:2002ad}. 
Our choice of setting the new parameters $\beta_{13},\ldots,\beta_{16}$
to zero does not change the results with respect to the $SU(3)$ case, since
we kept $\beta_{0}$ explicitly. If, on the other hand, we would have preferred to 
absorb $\beta_{0}$, the equivalence of both schemes could have only been restored
by taking non-vanishing values for $\beta_{13},\ldots,\beta_{16}$ in the $U(3)$ Lagrangian
as given by the Cayley-Hamilton identity.

There are, however, small differences between the $SU(3)$ and $U(3)$ results, 
if the same set of parameters is employed, and they are 
most easily seen in the positions of the poles
in \tabref{tab:ResPoleSU3} and \tabref{tab:ResPoleU3}
which change by up to $30\MeV$. 
These changes give a measure for the importance of the $\eta'$ contributions
within this approach. In the $SU(3)$ framework some effects of the
$\eta'$  are hidden in the coupling constants
of the fourth order chiral Lagrangian
\cite{Beisert:2001qb,Ecker:1989te,Herrera-Siklody:1998cr}, 
whereas
in the $U(3)$ theory the $\eta'$ is treated as a dynamical degree of freedom.
Hence, in order to reproduce the $SU(3)$ results more accurately, the coupling
constants would have to be modified slightly compensating the  $\eta'$ contributions
in the $U(3)$ framework.

\subsection{Comparison with previous work}

It is instructive to compare the present investigation with previous
work on coupled meson channels.
In \cite{Oller:1997ti} the second order $SU(3)$ Lagrangian 
was sufficient to reproduce the measured scattering
data below $1\GeV$. 
This work was done in cut-off regularization and 
with a reduced set of channels. 
When setting all fourth order couplings to zero in our 
approach we obtain very similar scattering data,
with or without the $\eta'$ channels,
which shows that at leading order the $\eta'$ has 
hardly any effect on low energy physics.

In \cite{Oller:1999hw}, for example, the approach was extended by including
the fourth order Lagrangian and a full
analysis of the scalar and vector channels in $SU(3)$ ChPT was performed. 
By adjusting the LECs the authors were able to obtain good agreement for all
presented data. 
The main difference to our scheme is
the expansion of the scattering amplitude. 
For the amplitude $A$ at a vertex we use the sum of the
second and fourth order amplitudes $A=A_2+A_4$, whereas in \cite{Oller:1999hw}
the inverse amplitude (IAM) expansion
\begin{equation}\label{eq:ResOOPAmp}
A=A_2(A_2-A_4)^{-1}A_2=A_2+A_4+A_4A_2^{-1}A_4+\ldots
\end{equation}
was used which is equal in fourth order. 
Both approaches differ significantly at energies 
well above $E_{\mathrm{cm}}=1\GeV$.
This difference can be easily understood by investigating the asymptotic 
dependence of $A$ with respect to the energy squared $s$.
The amplitudes $A_2$ and $A_4$ are linear and quadratic in $s$, respectively. 
While in our scheme the introduction of fourth order couplings 
increases the asymptotic power of $A$ to two, 
it is decreased to zero in the other scheme.

In a couple of papers on this subject the results
were refined. In the work \cite{GomezNicola:2001as} a full IAM analysis 
with manifest regularization independence, but without manifest 
unitarity is performed. The main advantage of this 
approach is the direct compatibility with chiral perturbation 
theory from which the one-loop amplitude is taken.
Here, the $K\pi$, $I=\sfrac{1}{2}$, scattering phase 
agrees with the experimental data. This is possibly due to
the $t$ and $u$ channel loops that are included in
the full IAM analysis. When they are dropped as in 
\cite{Oller:1999hw} or our analysis the scattering phase
is increased. The effect of those loops can be
simulated by a change of renormalization scale,
this is what we did by lowering $\mu$ to $0.8\GeV$ in 
this particular channel. 

Finally, we would like to comment on a possible extension to $SU(3)$ or $U(3)$ 
of the $SU(2)$ analysis described in \cite{Nieves:2000bx} where
emphasis was put on renormalization. 
For each channel a separate counterterm polynomial was introduced
to account for the infinities of the loop integral.  
The success of this method relies on the fact that
there are only three channels in $SU(2)$, $\pi\pi$ with
$(I,J)$ equals $(0,0)$, $(1,1)$ and $(2,0)$. 
In $SU(3)$ or $U(3)$ ChPT, however,
more than ten channels exist and each of them would require different 
coefficients for the polynomials. 
With such a large number of coefficients agreement with experimental data is
easily achieved without constraining most of the parameters
and the method would lose its predictive power.

\section{Conclusions}

In this work we have analyzed meson-meson scattering from
the $SU(3)$ and $U(3)$ chiral effective
Lagrangians in the $s$-wave channel by means of a coupled channel
Bethe-Salpeter equation.
We have presented the Bethe-Salpeter equation and solved it 
for a local interaction kernel.
Resonances are identified by relating them to poles in the 
analytical continuation of the scattering cross section and
multiplets of composed
states of two fundamental pseudoscalar mesons, i.e. $(\pi,K,\eta,\eta')$, are discussed.

We first investigated the $SU(3)$ case.
The fourth order Lagrangian was simplified by
taking only the most relevant parameters according to the OZI rule into account,
which are
$\cbeta{0}{0}$, $\cbeta{3}{0}$, $\cbeta{5}{0}$ and $\cbeta{8}{0}$.
In the isospin $I=0$ and $I=\sfrac{1}{2}$ channels we were able to 
fit the scattering phases up to about $1.2$-$1.3\GeV$, in analogy 
to results found, e.g., in \cite{Oller:1999hw}. 
Above $1.2\GeV$ deviations from the experimental phase shifts are observed as expected 
due to the omission of higher particle states, 
e.g., the $4\pi$ channel should become important in the
$I=0$ channel at these energies.
For the $I=\sfrac{3}{2}$ and $I=2$ channels reasonable agreement with
experiment is achieved  for center-of mass energies up to $1.5\GeV$
and no significant increase of the phase shifts is observed.

In a second step, the analysis was extended to $U(3)$ ChPT by including the
$\eta'$ explicitly.
Employing the {\it same} choice for the LECs as in the $SU(3)$ case and
neglecting new couplings of the $\eta'$ which are also OZI suppressed
(more generally: the 1/$N_c$ suppressed couplings) 
the results in this energy region were not altered considerably
and again the spectrum could be reproduced. This is a non-trivial
statement, since the coupling between the $\eta'$ channels with the other ones
may have destroyed the agreement of the pure $SU(3)$ case, and is in 
contradistinction to the results recently obtained in the meson-baryon sector \cite{Bass:2001np}.
Nevertheless, small effects from the inclusion of the 
$\eta'$ are observed which would require a slight readjustment of the
coupling constants, in order to reproduce the results of the $SU(3)$ case. 
In our approach and with our choice of the parameter values
the inclusion of the $\eta'$ does not yield
new resonances below $1.5\GeV$ that could be interpreted as quasi-bound states
of the $\eta'$ with a Goldstone boson.

We should mention that our fit to the phase shifts is not unique.
The OZI violating parameters which we have neglected here do not necessarily
need to be small and can contribute to meson-meson scattering. 
However, a small variation of these parameters could always be compensated
by small variations of 
$\cbeta{0}{0}$, $\cbeta{3}{0}$, $\cbeta{5}{0}$ and $\cbeta{8}{0}$.
The choice of the parameters made in the present investigation is in so far appealing
as it takes only a minimal set of four LECs into account while setting the remaining 
OZI violated couplings to zero.
Further phenomenological input such as the three pion decays of the $\eta$ and $\eta'$
\cite{Beisert:2003zs} may help to extract the values of some of the LECs more precisely
and clarify if this simplifying assumption for the LECs was justified.

\section*{Acknowledgements}
We are grateful to E. Marco for useful discussions. 
This work was supported in part by the Deutsche Forschungsgemeinschaft.


\bibliography{scalars} 

\ifx\publ\publj
\bibliographystyle{h-physrev4}
\else
\bibliographystyle{nb}
\fi

\ifx\publ\publj
\newpage
\section*{Tables}

\tabone
\tabtwo

\newpage
\fi

\section*{Figures}

\ifx\publ\publj
\figone
\figtwo
\newpage
\figthree
\figfour
\fi

\figfive
\figsix


\end{document}